\documentclass[12pt,a4paper,twoside]{article}
\usepackage{epsfig}
\usepackage{baltlat1}
\usepackage{wrapfig}
\def\araa{{\em ARAA}}

\def\apj{{\em ApJ}}
\def\apjs{{\em ApJS}}

\def\mnras{{\em MNRAS}}

\def\prl{{\em Phys.\ Rev.\ Lett.}}

\begin{document}
\ \
\vspace{0.5mm}

\setcounter{page}{1}
\vspace{8mm}

\titlehead{Baltic Astronomy, vol.~12, XXX--XXX, 2003.}

\titleb{Gravoturbulent Fragmentation}

\begin{authorl}
\authorb{Ralf S.\ Klessen}{1} and
\authorb{Javier Ballesteros-Paredes}{2}
\end{authorl}

\begin{addressl}
\addressb{1}{Astrophysikalisches Institut Potsdam, An der Sternwarte 16,
 14482 Potsdam, Germany}

\addressb{2}{Centro de Radioastronom\'ia y Astrof\'isica,
UNAM. Apdo. Postal 72-3 (Xangari), Morelia, Michoc\'an 58089, M\'exico}
\end{addressl}

\submitb{Received September XX, 2003}

\begin{abstract}
  
  We discuss star formation in the turbulent interstellar medium. We
  argue that morphological appearance and dynamical evolution of the
  gas is primarily determined by supersonic turbulence, and that stars
  form via a process we call gravoturbulent fragmentation. Turbulence
  that is dominated by large-scale shocks or is free to decay leads to
  an efficient, clustered, and synchronized mode of star formation. On
  the other hand, when turbulence carries most of its energy on very
  small scales star formation is inefficient and biased towards single
  objects.  
  
  The fact that Galactic molecular clouds are highly filamentary can
  be explained by a combination of compressional flows and shear. Some
  filaments may accumulate sufficient mass and density to become
  gravitationally unstable and form stars. This is observed in the
  Taurus molecular cloud. Timescales and spatial distribution of
  protostars are well explained by the linear theory of gravitational
  fragmentation of filaments.  The dynamical evolution, especially at
  late times, and the final mass distribution strongly depend on the
  global properties of the turbulence. In dense embedded clusters
  mutual protostellar interactions and competition for the available
  mass reservoir lead to considerable stochastic variations between
  the mass growth histories of individual stars.
\end{abstract}
\begin{keywords}
ISM: clouds, ISM: kinematics, stars: formation, turbulence
\end{keywords}


\sectionb{1}{Introduction}
\label{sec:intro}
Stars form by gravoturbulent fragmentation in interstellar clouds.
The supersonic turbulence ubiquitously observed in molecular gas
generates strong density fluctuations with gravity taking over in the
densest and most massive regions.  Once gas clumps become
gravitationally unstable, collapse sets in. The central density
increases until a protostellar object forms and grows in mass via
accretion from the infalling envelope. Various aspects of this process
have been discussed, e.g., by Hunter \& Fleck (1982), Elmegreen
(1993), Padoan (1995), Ballesteros-Paredes et al.\ (1999ab, 2003),
Klessen, Heitsch, \& Mac~Low (2000) or Padoan \& Nordlund (1999,
2002). See the reviews by Mac Low \& Klessen (2003) and Larson (2003).

In this proceedings paper we discuss the complex interplay of supersonic
turbulence and self-gravity and introduce the concept of gravoturbulent
fragmentation.  We argue that in typical star foming clouds turbulence
generates the density structure in the first place and then gravity takes over
in the densest and most massive regions.  In Section 2 we focus on spatial
distribution and timescale of star formation, then in Section 3, we discuss a
specific example of a star forming filament similar to those observed in
Taurus, and in Section 4 we speculate about the mass spectra of clumps and
stars in the context of the gravoturbulent fragmentation model.

\sectionb{2}{Spatial Distribution and Timescale of Star Formation}
\label{sec:location-time}
Supersonic turbulence plays a dual role in star formation. While it
usually is strong enough to counterbalance gravity on global scales it
will usually provoke collapse locally (e.g.\ Sasao 1973 for an early
analysis; or V{\' a}zquez-Semadeni et al.\ 2000; Mac~Low \& Klessen
2003; and Larson 2003 for recent reviews).  Turbulence establishes a
complex network of interacting shocks, where regions of high-density
build up at the stagnation points of convergent flows.  These gas
clumps can be dense and massive enough to become gravitationally
unstable and collapse when the local Jeans length becomes smaller than
the size of the fluctuation.  However, the fluctuations in turbulent
velocity fields are highly transient.  They can disperse again once
the converging flow fades away (Vazquez-Semadeni, Shadmehri, \&
Ballesteros-Paredes 2002).  Even clumps that are strongly dominated by
gravity may get disrupted by the passage of a new shock front (Klein, McKee \& Colella 1994, Mac Low et al.\ 1994).

For local collapse to result in the formation of stars, Jeans unstable,
shock-generated, density fluctuations therefore must collapse to sufficiently
high densities on time scales shorter than the typical time interval between
two successive shock passages.  Only then do they `decouple' from the ambient
flow pattern and survive subsequent shock interactions.  The shorter the time
between shock passages, the less likely these fluctuations are to survive. The
overall efficiency of star formation depends strongly on the wavelength and
strength of the driving source (Klessen et al.\ 2000, Heitsch, Mac~Low, \&
Klessen 2001) which both regulate the amount of gas available for collapse on
the sonic scale where turbulence turns from supersonic to subsonic
(V\'azquez-Semadeni, Ballesteros-Paredes, \& Klessen 2003).

The velocity field of long-wavelength turbulence is dominated by
large-scale shocks which are very efficient in sweeping up molecular
cloud material, thus creating massive coherent structures. These
exceed the critical mass for gravitational collapse by far, because
the velocity dispersion within the shock compressed region is much
smaller than in the ambient turbulent flow. The situation is similar
to localized tur\-bulent decay, and quickly a cluster of protostellar
cores builds up. Both decaying and large-scale turbulence lead to a
{\em clustered} mode of star formation. Prominent examples are the
Trapezium Cluster in Orion with a few thousand young stars, but also
the Taurus star forming region which is historically considered as a
case of isolated stellar birth. Its stars, however, have formed almost
simultaneously within several coherent filaments which apparently are
created by external compression (see Ballesteros-Paredes et al.\ 
1999). This renders it a clustered star forming region in the sense
of the above definition.

The efficiency of turbulent fragmentation is reduced if the driving
wavelength decreases. There is less mass at the sonic scale and the
network of interacting shocks is very tightly knit. Protostellar cores
form independently of each other at random locations throughout the
cloud and at random times. There are no coherent structures with
multiple Jeans masses. Individual shock generated clumps are of low
mass and the time interval between two shock passages through the same
point in space is small.  Hence, collapsing cores are easily destroyed
again. Altogether star formation is inefficient. This scenario then
corresponds to an {\em isolated} mode of star formation. Stars that
truely form in isolation are, however, very rarely observed -- most
young stars are observed in clusters or at most loose aggregates. From
a theoretical point of view, there is no fundamental dichotomy between
these two modes of star formation, they rather define the extreme ends
in the continuous spectrum of the properties of turbulent molecular
cloud fragmentation.

Altogether, we call this intricate interaction between turbulence on the one
side and gravity on the other -- which eventually leads to the transformation of
some fraction of molecular cloud material into stars as described above -- {\em
  gravoturbulent fragmentation}. To give an example, we discuss in detail the
gravitational fragmentation in a shock-produced filaments that closely
resembles structures observed in the Taurus star forming region.

\sectionb{3}{Gravitational Fragmentation of a Filament in a Turbulent Flow}\label{Taurus:sec} 

In Taurus, large-scale turbulence is thought to be responsible for the
formation of a strongly filamentary structure (e.g.\
Ballesteros-Paredes et al.\ 1999). Gravity within the filaments should then be
considered as  the main mechanism for forming cores and stars. Following earlier ideas
by Larson (1985), Hartmann (2002) has shown that the Jeans length within a
filament, and the timescale for it to fragment are given by
\begin{eqnarray}
\lambda_J  &= & 1.5 \ T_{10}\ A_V^{-1} \ {\rm pc,} \label{lambda-lee:eq} \\
\tau  & \sim  & \ 3.7 \ T^{1/2}_{10}\ A_V^{-1} \ {\rm Myr.}
        \label{timescale-filament:eq} 
\end{eqnarray}
where $T_{10}$ is the temperature in units of 10$\,$K, and $A_V$ is the
visual extinction through the center of the filament. 
By using a mean visual extinction for starless cores of $A_V\sim 5$
(Onishi et al.\ 1998), equation 1 gives a characteristic
Jeans length of $\lambda_J\sim 0.3\,$pc, and collapse should occur in
about 0.74$\,$Myr. Indeed, Hartmann (2002) finds $3-4$
young stellar objects per parsec with agrees well with the above
numbers from linear
theory of gravitational fragmentation of filaments.

In order to test these ideas, we resort to numerical simulations. We analyze a
SPH calculation (Benz 1990, Monaghan 1992) of a star forming region that was
specifically geared to the Taurus cloud. Details on the numerical
implementation, on performance and convergence properties of the method, and
tests against analytic models and other numerical schemes in the context of
turbulent supersonic astrophysical flows can be found in Mac~Low et al.\ 
(1998), Klessen \& Burkert (2000, 2001) and Klessen et al.\ (2000).

This simulation has been performed without gravity until a particular, well
defined elongated structure is formed. We then turn-on self-gravity. This
leads to localized collapse and a sparse cluster of protostellar cores builds
up. Timescale and spatial distribution are in good agreement with the Hartmann
(2002) findings in Taurus. For illustration, we show eight column density
frames of the simulation in figure 1. The first frame shows
the structure just before self-gravity is turned-on, and we note that the
filament forms cores in a fraction of Myr. The timestep between frames is
0.1$\,$Myr. The mean surface density for the filament is
0.033$\,$g$\,$cm$^{-2}$, corresponding to a visual extinction of $\sim$7.5.
Using equations 1 and 2 this value
gives a Jeans length of $\lambda_J \sim 0.2\,$pc, and a collapsing timescale
of $\tau \sim 0.5\,$Myr. Note from figure 1 that the first
cores appear roughly at $\tau \sim 0.3\,$Myr, although the final structure of
collapsed objects is clearly defined at $t=0.5\,$Myr. The typical
separation between protostellar cores (black dots in figure 1) is about the Jeans length
$\lambda_J$.

\begin{figure}
\centerline{\psfig{figure=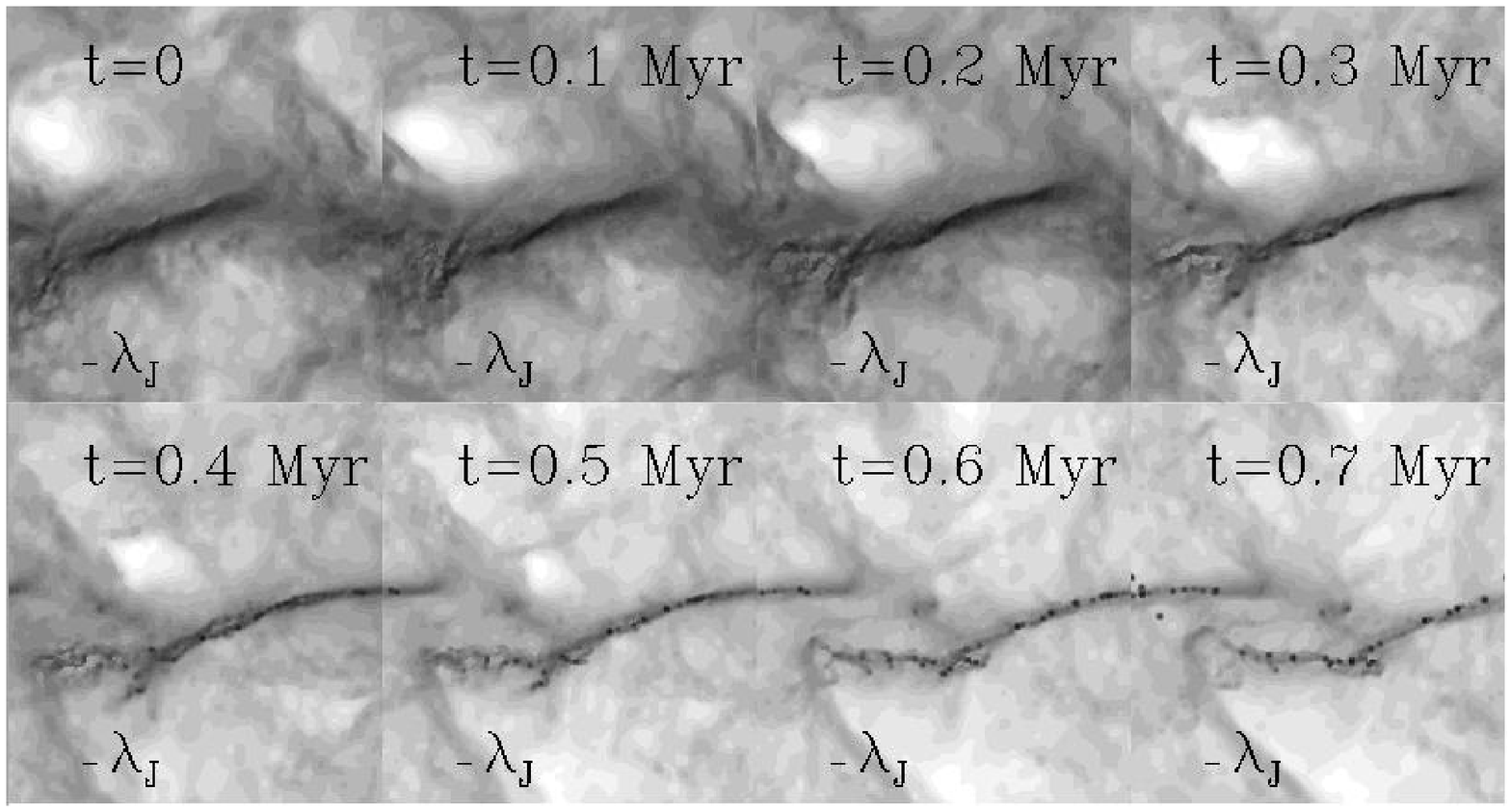,width=120truemm, angle=0}}
\captionb{1}{%
  Evolution of the column density of an SPH simulation. The filament
  in the first frame (before self-gravity is turned-on) shows that
  turbulence is responsible in forming this kind of structures. The
  small bar in the bottom-left of each frame denotes the Jeans length
  (equation 1) at this time. At later times, self-gravity is turned on
  and the filament suffers gravitational fragmentation on a free-fall
  timescale (equation 2).}
\end{figure}

This example demonstrates that indeed turbulence is able to produce a strongly
filamentary structure and that at some point gravity takes over to form
collapsing objects, the protostars. However, the situation is quite complex.
Just like in Taurus, the filament in figure 1 is not a perfect cylinder, the
collapsed objects are not perfectly equally spaced as predicted by idealized
theory, and protostars do not form simultaneously but during a range of times
(between $t \approx 0.3$ and 0.6$\,$Myr). Even though the theory of
gravitational fragmentation of a cylinder appears roughly, it becomes clear
that the properties of the star forming region not only depend on the
conditions set initially but are influenced by the large-scale turbulent flow
during the entire evolution. Gravoturbulent fragmentation is a continuous
process that shapes the accretion history of each protostar in a stochastic
manner (e.g.\ Klessen 2001a).

\sectionb{4}{Mass Spectra of Clumps and Protostellar Cores}
\label{sec:mass-spectra}
The dominant parameter determining stellar evolution is the mass. We discuss now how the final stellar masses may depend on the gravoturbulent
fragmentation process, and analyze four  numerical
models which span the full parameter range from strongly clustered to very
isolated star formation (for full detail see Klessen 2001b).

Figure 2 plots the mass distribution of all gas clumps, of the subset
of Jeans critical clumps, and of collapsed cores. We show four
different evolutionary phases, initially just when gravity is
`switched on', and after turbulent fragmentation has lead to the
accumulation of $M_{\rm \large *}\approx 5$\%, $M_{\rm \large
  *}\approx 30$\% and $M_{\rm \large *}\approx 60$\% of the total mass
in protostars.

\begin{figure}[th]
\unitlength1cm
\begin{center}
\begin{picture}(10.0,8.0)
\put( 0.0, -0.8) {\epsfxsize=10.0cm \epsfbox{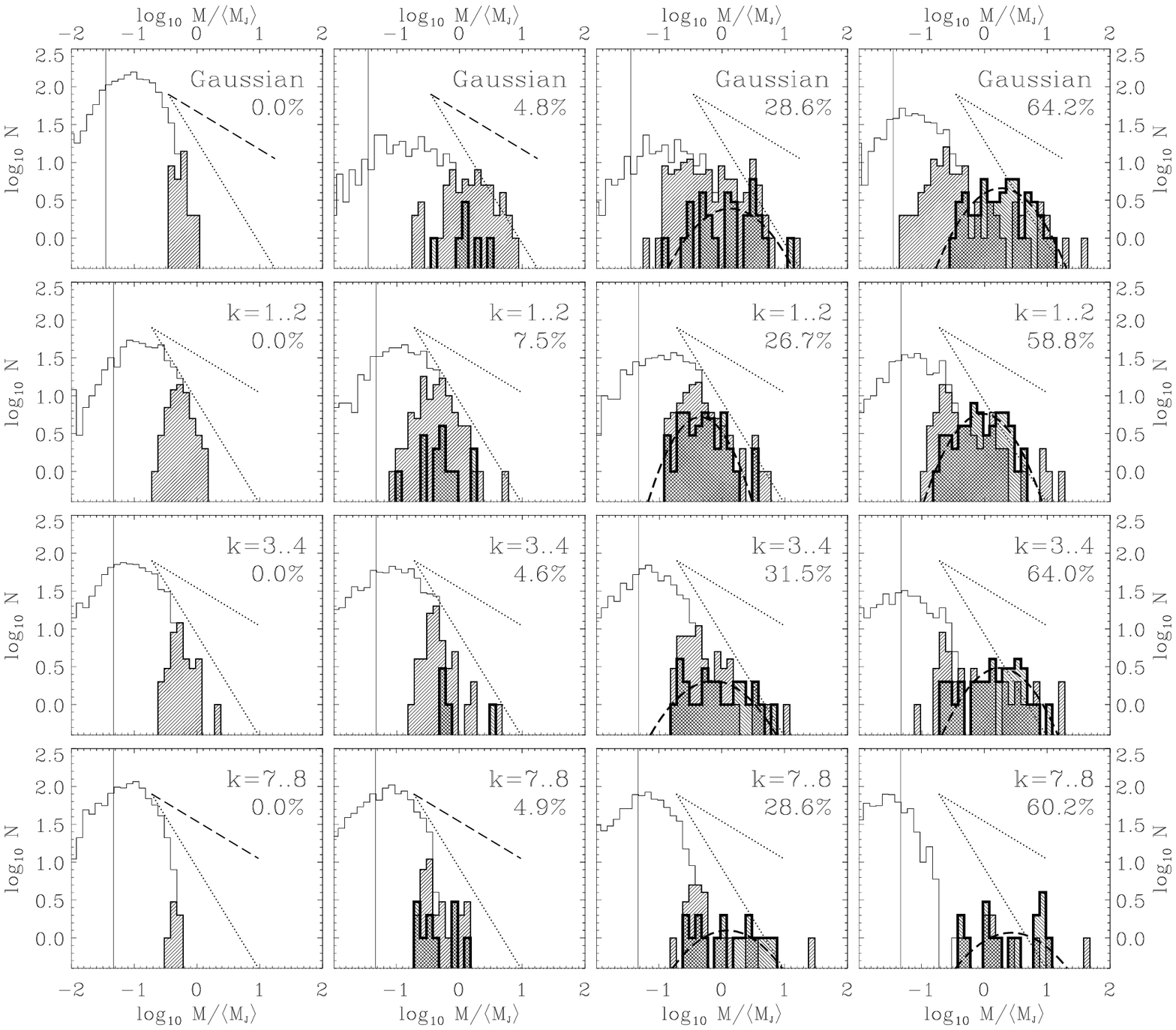}}
\end{picture}
\end{center}
\captionb{2}{Mass spectra of protostars (hatched thick-lined
  histograms), of gas clumps (thin lines), and of the subset of Jeans
  unstable clumps (thin lines, hatched distribution). Different 
  evolutionary phases are defined  by the fraction of mass converted into
  protostars and are indicated in the upper right corner of each plot. 
  Masses are
  binned logarithmically and normalized to the average Jeans mass
  $\langle M_{\rm J}\rangle$. (From Klessen 2001b.)
}
\end{figure}

In the  completely pre-stellar phase the clump mass spectrum
is  very steep (about Salpeter slope or less) at the high-mass end. It
has a break and
gets shallower below $M \approx 0.4 \,\langle M_{\rm J} \rangle$ with
slope $-1.5$. The spectrum strongly declines
beyond the SPH resolution limit. Individual clumps are
hardly more massive than a few $\langle M_{\rm J}\rangle$.

Taking gravitational evolution into account modifies the distribution
of clump masses considerably. As clumps merge and grow bigger, the
spectrum becomes flatter and extends towards larger masses.
Consequently the number of cores that exceed the Jeans limit
increases. This is most evident in the Gaussian model of decayed
turbulence, the clump mass spectrum exhibits a slope $-1.5$.


The mass spectrum depends on the wavelength of the dominant
velocity modes.  Small-scale turbulence does not allow for massive,
coherent and strongly-selfgravitating structures, and together with
the short interval between shock passages, this prohibits efficient
merging and the build up of a large number of massive clumps. Only few
clumps become Jeans unstable and collapse to form stars. This occurs  at random locations and times. The clump mass spectrum remains steep as in the
case without gravity.  Increasing the driving wavelength leads to more
coherent and rapid core formation, leading to a larger number of
cores.

Long-wavelength turbulence or turbulent decay produces a core mass
spectrum that is well approximated by a {\em log-normal}. It roughly
peaks at the {\em average thermal Jeans mass} $\langle M_{\rm
  J}\rangle$ of the system (see Klessen \& Burkert 2000, 2001) and is
comparable in width with the observed IMF (Kroupa 2002). 
The log-normal shape of the mass distribution may be
explained by invoking the central limit theorem (e.g.\ Zinnecker
1984), as protostellar cores form and evolve through a sequence of
highly stochastic events (resulting from supersonic turbulence and/or
competitive accretion).

ACKNOWLEDGEMENTS.\  We thank Lee Hartmann, Mordecai-Mark Mac~Low,  and Enrique
V\'azquez-Semadeni for many stimulating discussions and fruitful
collaboration. RSK acknowledges support by the Emmy Noether Program of
the Deutsche Forschungsgemeinschaft (grant no.\ KL1358/1), and JBP thanks for  
 support from Conacyt's grant I39318-E.
\goodbreak

\References
{\footnotesize

\def\ref{{}}

  \noindent  Adams, F.\ C., Myers, P.\ C., 2001, \apj, 553, 744

  \noindent  Benz, W., 1990, in The Numerical Modeling of Nonlinear
  Stellar Pulsations, ed.\ J.\ R.\ Buchler (Dordrecht: Kluwer), 269

  \noindent  Ballesteros-Paredes, J., Hartmann, L.,  V{\'
    a}zquez-Semadeni, E., 1999a, \apj, 527, 285

  \noindent  Ballesteros-Paredes, J., V{\' a}zquez-Semadeni, E., 
  Scalo, J., 1999b, \apj, 515, 286

  \noindent  Elmegreen, B.\ G., 1993, \apj, 419, L29

  \noindent Hartmann, L.\ 2002, \apj,  578, 914

  \noindent  Heitsch, F., Mac~Low, M.-M., Klessen, R.~S., 2001, \apj, 547, 280

  \noindent  Hunter, J.~H.~\& Fleck, R.~C.\ 1982, \apj, 256, 505

  \noindent  Klein, R.\ I.,  McKee, C.\ F., Colella, P., 1994,
    \apj, 420, 213

  \noindent  Klessen, R.~S., 1997, \mnras, 292, 11

  \noindent  Klessen, R.~S., 2001a, \apj, 550, L77

  \noindent  Klessen, R.~S., 2001b, \apj, 556, 837

  \noindent  Klessen, R.~S., Burkert, A., 2000, \apjs, 128, 287

  \noindent  Klessen, R.~S., Burkert, A., 2001, \apj, 549, 386

  \noindent  Klessen, R.~S., Heitsch, F., Mac~Low, M.-M., 2000, \apj,
    535, 887

  \noindent  Larson, R.~B.\ 1985  \mnras, 214, 379

  \noindent  Kroupa, P.\ 2002, Science, 295, 82

  \noindent  Larson, R.~B.\ 2003, {Rep.\ Prog.\
    Phys.}, 66, 1651

  \noindent  Mac~Low, M.-M., \& Klessen,
  R.\ S. 2003, Rev.\ Mod.\ Phys., 76, 124

  \noindent  Mac~Low, M.-M., Klessen, R.\ S., Burkert, A., Smith, M.\ 
  D., 1998, \prl, 80, 2754

  \noindent  Mac Low, M.-M.,McKee, C.\ F., Klein, R.\ I., Stone, J.\ 
    M., Norman, M.\ L., 1994, \apj, 433, 757

  \noindent  Monaghan, J.~J. 1992, \araa, 30, 543

  \noindent  Padoan, P., 1995, \mnras, 277, 337

  \noindent  Padoan, P., Nordlund, \AA., 1999, \apj, 526, 279

  \noindent  Padoan, P.,  Nordlund, {\AA}., 2002, \apj, 576, 870

  \noindent  V{\' a}zquez-Semadeni, E., Ballesteros-Paredes, J.,
  Klessen, R.~S., 2003, \apj, 585, L131

  \noindent  V{\' a}zquez-Semadeni, E., Shadmehri, M., Ballesteros-Paredes, J.,
  2002, \apj, submitted (astro-ph/0208245)

  \noindent  Zinnecker, H., 1984, \mnras, 210, 43

}

\end{document}